\documentclass[12pt,preprint]{aastex}

\newcommand\degree  {\ifmmode{^\circ}\else{$^{\circ}$}\fi}
\newcommand\amin   {\ifmmode{^\prime}\else{$^\prime$}\fi}
\newcommand\pamin  {\ifmmode{{\rlap.}^\prime}\else{${\rlap.}^\prime$}\fi}
\newcommand\asec   {\ifmmode{^{\prime \prime}}\else{$^{\prime \prime}$}\fi}
\newcommand\pasec  {\ifmmode{{\rlap.}^{\prime \prime}}\else{${\rlap.}^{\prime \prime}$}\fi}

\newcommand\kms    {~km~s$^{-1}$}

\newcommand\lax    {\ifmmode{_<\atop^{\sim}}\else{${_<\atop^{\sim}}$}\fi}
\newcommand\gax    {\ifmmode{_>\atop^{\sim}}\else{${_>\atop^{\sim}}$}\fi}
\def\mic.{$\mu$m}
\def\kms {\hbox{${\rm km\, s}^{-1}$}}

\shorttitle{Water Masers Toward Ultracompact HII Regions}
\shortauthors{Kurtz and Hofner}

\begin{document}

\title{Water Masers Toward Ultracompact HII Regions}

\author{S. Kurtz}
\affil{Centro de Radioastronom\'\i a y Astrof\'\i sica, UNAM,
  Apartado Postal 3-72, 58089, Morelia, Michoac\'an, M\'exico}
\email{s.kurtz@astrosmo.unam.mx}
\and
\author{P. Hofner}
\affil{ Physics Department, New Mexico Institute of Technology, Socorro, 
NM 87801; and National Radio Astronomy Observatory, P.O. Box O, Socorro, 
NM 87801}
\email{phofner@nrao.edu}

\begin{abstract}
We present a survey in the $6_{16}-5_{23}$ rotational H$_2$O transition
toward 33 galactic ultracompact HII regions.  Maser emission is detected
toward 18 of these sources; two are new detections.  High quality spectra 
are provided for all 18 sources.   We discuss the detection rate of this
survey and the correlation of various maser properties with other physical
parameters.
In addition, we report wide-bandwidth (316~km~s$^{-1}$), moderate-resolution 
($\sim 3''$)
H$_2$O maser observations of the HH80-81 region.  We report the first 
detection of water maser emission at the approximate velocity of the
molecular core.  This emission is coincident with the extreme tip of
the thermal jet, and well-removed from the much stronger and well-known
maser emission at the position of VLA-3.
\end{abstract}


\section{Introduction}

Since their discovery over 30 years ago, water masers have been found
in a variety of astronomical settings, primarily, but not exclusively,
in the environs of evolved stars and of star-forming regions.
Extensive surveys and classifications of water masers have been made
(for the northern hemisphere) by researchers at the Arcetri
Observatory, including Cesaroni et al. (1988), Comoretto et
al. (1990), Palagi et al. (1993), Brand et al. (1994), Valdettaro et
al. (2001), and Brand et al.  (2003). References to a number of
southern surveys may be found in Braz et al. (1989).  Additional
studies, focusing on particular classes of objects, include Wouterloot
\& Walmsley (1986), Churchwell, Walmsley \& Cesaroni (1990); Palla et
al. (1991, 1993); Felli, Palagi, \& Tofani (1992), 
Henning et al.  (1992); Palumbo et al. (1994);
Codella, Felli, \& Natale (1996, and references therein), Hofner and
Churchwell (1996), Furuya et al.  (2003), and de Gregorio-Monsalvo et
al. (2004).  The present work extends the study of water masers in a
well-defined sample of objects, in particular, water masers near
ultracompact (UC) HII regions.

Studies too numerous to cite have found that water masers can trace
circumstellar disks, molecular outflows, and ionized jets, all of
which may be present at some moment during the massive star formation
process.  In some star-forming regions water masers are found in close
proximity (often in projection against) the ionized gas of HII regions
(e.g., Hofner \& Churchwell 1996).  However, in most cases, water masers seem
to be physically associated with warm molecular gas, avoiding the
HII region (e.g., Cesaroni et al.  1994).  Trends in the coincidence of
water, methanol, and hydroxyl masers have been interpreted to reflect
an evolutionary sequence in the star formation process (see the
discussions in Garay \& Lizano 1999 and Beuther et al. 2002).

Despite the abundance of observational data and significant theoretical
understanding of the pumping mechanisms (e.g., Kylafis \& Norman 1991, 
and Elitzur, Hollenbach \& McKee 1989), water masers continue to 
challenge our understanding, precisely because of the multiple 
phenomena that they trace.  Further clarification of the physical role
of water masers in the proximity of UC HII regions is needed, and the
present work is intended to contribute to this topic.

In this paper we present the results of a single-dish survey of 33 UC
HII regions and an interferometric study of one of these --- HH80-81,
also known as GGD 27 and G10.84$-$2.59.  In \S\S 2 and 3 we describe
the single-dish observations and discuss their results.  In \S 4 we
present the interferometric observations and discuss their results.
We summarize our findings in \S 5.

\section{SINGLE-DISH OBSERVATIONS AND DATA REDUCTION}

Observations of the $6_{16}-5_{23}$ rotational transition of 
H$_2$O were obtained on  1995 September  22 -- 24 with the 100~m
telescope of the Max Planck Institute for Radioastronomy near
Effelsberg, Germany. 
We used the facility 1.3~cm receiver and the 1024
channel autocorrelator. System temperatures ranged from 140
to 280~K. We observed with bandwidths of 25 and 12.5~MHz,
which provide resolutions of 0.33 and 0.17~km~s$^{-1}$
respectively.
The half-power beamwidth, measured through cross-scans on strong
point sources, was about $40''$.
Flux calibration was based on observations of NGC$\,$7027, for
which we assumed a flux density of 5.86~Jy. An elevation
dependent gain correction was applied to the data.
We observed the strong source G34.26+0.15
daily to monitor the influence of varying weather conditions on the
calibration. From these observations we conclude that the calibration 
accuracy is better than $25\,$\%.
Pointing corrections were obtained every hour from cross-scans of 
strong point sources at small angular distances from the program 
sources; typical corrections were $5''$.

Our candidate sources were taken from Kurtz, Churchwell \& Wood 1994
(hereafter KCW). Table~1 lists the observed sources in column 1.
Columns 2 and 3 give the pointing position in equatorial coordinates,
column 4 indicates the range of observed LSR velocities, and column 5
reports our detection limits. The median 3$\sigma$ detection limit was
0.34 Jy. In column 6 we indicate whether water masers were detected.
 
The spectra were baseline subtracted, summed, and smoothed to a final
resolution of 0.33~km~s$^{-1}$.
In Table~2 we report the observed parameters of the detected masers. 
In column~1 we list the source name, while columns 2 and 3 give
the flux density and LSR velocity of the strongest feature in the
spectrum. In column~4 we list the velocity range over which emission
in the H$_2$O line was detected and in column~5 the 
velocity-integrated flux density of the strongest feature.

\section{SINGLE-DISH RESULTS}

Maser emission was detected in 18 of the 33 sources observed; a 55\%
detection rate.  Two of these, G11.11$-$0.40 and G18.30$-$0.39 are new
detections, not previously reported in the literature. The spectra of
all 18 sources are shown in Figure~1.

\subsection{Variability}

Variability of water masers is well-known and our observations
confirm this in a number of cases.  Four sources show
particularly strong variations. G10.84$-$2.59 shows
a flux density of 205 Jy in our observations. It is reported at
8~Jy by Mart\'\i{}, Rodr\'\i guez \& Torrelles (1999), and 53~Jy
by Cesaroni et al. (1988), while Codella et al. (1995) report a null
detection with an upper limit of 3.6~Jy.  We detect a 4~Jy maser in
the G35.02+0.35 field, for which Cesaroni et al. (1988) report 63~Jy.
G80.87+0.42 is undetected by us and by Palla et al. (1991) but a
106 Jy detection is reported by Han et al. (1998).  G106.8+5.31 we
detect with 15~Jy, while Comoretto et al. (1990) detected 27~Jy and
Cesaroni et al. (1988) report 250~Jy. 

The impact of variability on our detection rate is not clear.  Of the
15 null detections we report, 13 (all but G9.88$-$0.75 and
G18.15$-$0.28) have observations reported in the literature.  Of the
13, only two (G80.87+0.42 and G111.28$-$0.66) have reported
detections.  Using a simple statistical argument, applicable to their
sample, Miralles, Rodr\'\i guez \& Scalise (1994) argue that essentially
all sources in their sample may show water maser emission at some point,
with null detections arising from variability.  The Miralles et al. 
sample was quite similar to our own in the sense that both are 
constituted of high mass star formation regions; hence it is possible
that most of the 15 null detections we report do harbor water masers.

\subsection{Maser Velocities}

In figure 2 we plot the maser line velocities versus associated
molecular line velocities.  Except for G10.84$-$2.59 and G106.80+5.31,
the molecular gas velocities are taken from the CS observations of
Bronfman, Nyman \& May (1996).  For the two sources mentioned, CS data
are not available; CO velocities are used instead (Kurtz 1993).  
As expected, the molecular line
velocities and the maser velocities are in generally good agreement,
with a median difference of 4.5~km~s$^{-1}$.  
The notable exception is
G10.84$-$2.59, also known as GGD~27 and HH~80-81, which shows a
$|V_{CO} - V_{H_2O}|$ of $\sim$85~km~s$^{-1}$ for the peak maser
component.  The large blueshift of the peak maser component with
respect to the ambient cloud was reported by G\'omez, Rodr\'\i guez \&
Mart\'\i{}, (1995), who suggest that outflow activity may be
occurring.  Our spectrum shows maser emission ranging from $-$80.6 to
+4.2~km~s$^{-1}$.  Though relatively weak, the most redshifted maser
component we detect differs from the cloud velocity by only
6.8~km~s$^{-1}$ (see \S4.1).  Prior high resolution
(interferometric) observations of this region used relatively narrow
bandpasses, centered near the strongest component at approximately
$-70$~km~s$^{-1}$, and thus did not detect the more redshifted
components.  To locate these components with respect to the
blueshifted masers and the free-free emission sources in the field, we
made moderate-resolution interferometric observations that we report
in \S 4.

The velocity range of maser emission is also shown in figure 2.  Two
sources in our sample, G10.84$-$2.59 and G69.54$-$0.89, presented
particularly broad ranges in maser velocities.  For G10.84 the total
range is 85 km~s$^{-1}$ (or 100 km~s$^{-1}$ including the VLA
detection of \S4) while for G69.54$-$0.89 the range was 116
km~s$^{-1}$. We are unaware of broadband interferometric observations
of G69.54$-$0.89 (ON-1), which would include the full 116 kms~s$^{-1}$
range.  Such observations would be worthwhile, particularly to locate
the extreme velocity components with respect to other star formation
signposts in the field (e.g., Kumar, Tafalla \& Bachiller 2004).

\subsection{Correlations}

No correlation was found between the free-free radio continuum flux
densities and the peak maser component flux densities.  No systematic
trends with respect to the far-infrared (FIR) luminosity (taken from
KCW) were seen, either in the detection/non-detection of masers, in
the flux density of the strongest maser component, or in the velocity
range over which maser emission was detected.  Likewise the range of
observed maser velocities showed no correlation with the free-free
flux density, the strength of the peak maser component, the CS (or CO)
linewidth, or {\it IRAS} colors.

We do find a trend for increasing maser luminosity with increasing
far infrared luminosity.  A linear fit provides results that are
consistent, within the uncertainty, with the results of Wouterloot \&
Walmsley (1986), Palagi et al. (1993), and  Miralles et al. (1994). 
Palagi et al. have a much larger sample and we consider their fit to 
be more reliable than our own.

Neither the null detections nor the detected sources showed evidence
for correlation with {\it IRAS} colors.  We note, however, that within
our sample there were a number of sources with large
F$_{60}$/F$_{25}$, F$_{60}$/F$_{12}$, and F$_{25}$/F$_{12}$
ratios (with respect to the sample median); without exception these
reddened sources showed maser emission.  Large F$_{100}$/F$_{60}$ ratios, 
however, are evenly mixed between detections and non-detections.
These trends are in general agreement with the results of Palla et al.
(1991) who found higher detection rates for redder sources.

\subsection{Analysis of the Detection Rate}

Numerous water maser surveys of star-forming regions have been made,
several of which used selection criteria similar to our own.  Three
such studies are particularly relevant here: Churchwell, Walmsley \&
Cesaroni (1990, hereafter CWC),  Palla et al. (1991, 1993), and 
Palumbo et al. (1994) and Codella et al. (1995).

The CWC source list was comprised of two parts: regions known to
contain UC~HII regions (64) and regions likely to contain UC~HII
regions (based on their {\it IRAS} colors), but not confirmed to
contain them (20).  They report a 67\% water maser detection rate for
the former group and a 65\% detection rate for the latter group.
Palla et al. (1991) found a 26\% detection rate for their ``high''
group, which had {\it IRAS} $60~\mu m$ fluxes greater than 100~Jy and
whose {\it IRAS} colors matched the selection criteria of Wood
\& Churchwell (1989, hereafter WC).  These color criteria require that
log (F$_{60}$/F$_{12}$) $\ge 1.30$ and log (F$_{25}$/F$_{12}$) $\ge
0.57$.  
Palumbo et al. (1994) and Codella et al. (1995) reviewed the
literature and performed additional observations of regions with {\it
  IRAS} colors indicative of star-forming molecular cloud cores.
Codella et al. identify 672 {\it IRAS} sources meeting the WC color
criteria and report a water maser detection rate of 20\%.

In the present work, we have used the same WC color criteria as the
other three surveys, and report a 55\% detection rate.  Thus, of four
samples meeting the WC color criteria, the reported water maser
detection rates are 67\% (CWC), 55\% (this paper), 26\% (Palla
et al. 1991) and 20\% (Codella et al. 1995).

As noted by Palla et al. (1991), sensitivity may explain part of the
differing detection rates.  Our typical $3\sigma$ detection level, as
that of CWC, is about a factor of 10 lower than the corresponding
detection level of Palla et al. (1991) and Codella et al. (1995).
Applying a 5~Jy cut-off level (typical of the detection levels of the
other surveys)lowers our detection rate to 36\% and that of CWC to
35\%.

In addition, the CWC sample, and our own, are biased toward brighter
infrared sources. Most of the CWC sources and all of our own have {\it
  IRAS} $100~\mu m$ flux densities greater than 1000~Jy.  All of
these sources have F$_{60~\mu m} > 100$~Jy, which is the flux cut-off
proposed by Codella, Felli \& Natale (1994) below which {\it IRAS}
sources cannot be reliably associated with UC~HII regions.  This same
flux limit, when imposed by Palla et al. (1993), more than halved
their maser detection rate.  Blind radio surveys, observing {\it IRAS}
sources with WC colors in search of UC~HII regions, suggest that the 
detection rate can fall from around 80\% for bright sources, with
$F_{100} > 1000$~Jy, to around 40\% for dimmer sources, with
$1000 \;{\rm Jy} > F_{100} > 100\; {\rm Jy}$ (Kurtz 1995).
It is plausible that the difference of 10-15\% in the detection rates,
after allowing for sensitivity effects, might result from a lower
success rate of the color selection criteria for weaker sources.

A reasonable interpretation is that the CWC detection rate, and the
rate we report here, are representative of the fractional lifetime
during which massive star formation regions with UC~HII regions
exhibit detectable (i.e., allowing for variability) maser emission.
On the other hand, the relatively low rates reported by Palla et al.
(1991, 1993) and by Codella et al.  (1995) probably reflect their
lower sensitivity and imperfections in the WC color selection criteria
for weak {\it IRAS} sources.  I.e., differing selection criteria and
differing sensitivities are the probable explanation for the distinctly
different detection rates.

Our sample is small and deliberately biased toward star-forming regions
that have already formed massive stars.  It is likely that water
masers occur in younger (proto-stellar) objects, but our results cannot 
be taken to represent isolated earlier evolutionary stages.  Rather, they
represent the typical case of massive star-forming regions where several
different evolutionary stages are present at the same time.

\section{ VLA OBSERVATIONS of HH80-81}

\subsection{Motivation and Observational Details}

Following the original water maser detection in the HH80-81 region
(G10.84$-$2.59; GGD~27) by Rodr\'\i guez et al. (1978), most observers
have concentrated on the so-called ``high velocity maser'', at
$V_{lsr} \approx -70$ \kms.  Our high sensitivity 100~m spectrum shows
evidence for maser emission at $V_{lsr} \approx$ +3~\kms, similar to
the molecular core velocity of +11 -- +13~\kms {}(Rodr\'\i guez et al. 1980,
G\'omez et al. 2003). The location of this ``positive velocity maser'' is
interesting, particularly because the high velocity maser is {\it not}
coincident with the thermal jet or its suspected driving source
(G\'omez, Rodr\'\i guez \& Mart\'\i{} 1995).  
To locate the positive velocity maser, we made
observations with the Very Large Array (VLA) of the NRAO\footnote{The
National Radio Astronomy Observatory is a facility of the National
Science Foundation operated under cooperative agreement by Associated
Universities, Inc.}  on 1999 June 3 and 1999 November 30.  The June
observations were at relatively low spatial and spectral resolution
but with broad velocity coverage.  The November observations offered
less velocity coverage, but at higher spatial and spectral resolution.
We describe the two observational periods below.

Observations were made on 1999 June 3 while the array was being
reconfigured from the D to the A configuration.  The two outermost
antennas on each arm were not available for observations.  Twenty
antennas, at D-array stations, were used, resulting in a resolution of
4\farcs 5 $\times$ 2\farcs 3.  Channel widths of 48.8~kHz provided a
velocity resolution of 0.66~\kms.  The data were edited and calibrated
using standard AIPS procedures.  Bandpass calibration was not applied
to the data, nor was self-calibration performed. The flux density
scale was set by observations of 3C286 with an adopted flux density of
2.518~Jy.  The phase calibrator was 1820$-$254, with a bootstrapped
flux density of 0.72~Jy.  We observed in five overlapping bands of 127
channels, centered at LSR velocities of $-$158, $-$92.5, $-$30, +32.5,
and +95 \kms.  The velocities were chosen so that adjacent bands had
16 channels of overlap which were discarded without loss of velocity
coverage.  Each velocity dataset was imaged and cleaned with the AIPS
task IMAGR, using uniform $uv$ weighting with a robustness of $-$0.2.
The cubes were initially cleaned to a flux level of 5 times the
theoretical $rms$ and inspected for maser emission.  Channels with
maser emission were then cleaned to a 2$\sigma$ level.  The resulting
image cubes were merged to produce a single 480 channel cube with a
velocity coverage of 316~\kms, from $-$190 to +126~\kms.

The 1999 June observations detected a positive velocity maser but had
inadequate spectral resolution to resolve the line.  The maser was
detected not at the +3~km~s$^{-1}$ of the 100-m detection, but rather
at a velocity of +20~km~s$^{-1}$.  To confirm the detection and to
provide better line parameters, we re-observed this spectral feature
with the VLA on 1999 November 30.  The array was in the
B-configuration (0\farcs 52 $\times$ 0\farcs 26 resolution), and we
observed for 14~min on-source, with a spectral resolution of 12.2~KHz
(0.16~km~s$^{-1}$) and 20~km~s$^{-1}$ bandwidth.  No flux calibrator
was observed, but inspection of the $uv$-data of the phase calibrator
(1820$-$254) suggests its flux density was 1.1~Jy, which is consistent
with values from VLA calibrator database during that period.  We adopt
this value, but caution that the absolute flux density calibration may
have an uncertainty as large as 20\%.  No bandpass calibration was
performed.  Imaging followed the same procedures as the 1999 June
observations.

\subsection{ Results: Maser Positions and Velocities}

Two sites of maser emission were detected by our broad-band
observations: the previously known, high-velocity maser coincident
with VLA-3 (G\'omez et al. 1995; Mart\'\i{} et al. 1999) and the newly
detected +20~km~s$^{-1}$ maser.  The maser at VLA-3 is substantially
the stronger of the two and contains multiple velocity features (see
figure 3).  The +20~km~s$^{-1}$ maser is much weaker and shows a
single velocity feature (see figure 4).  The VLA-3 maser is about
5$''$ to the east of the HH80-81 thermal jet axis, while the
20~km~s$^{-1}$ maser is aligned with the thermal jet (see figure~5).

The +20~km~s$^{-1}$ maser line profile was fully resolved by the 1999
November observations (see inset to figure~4).  The gaussian line FWHM
is 0.8~km~s$^{-1}$ with a central velocity of 19.9~km~s$^{-1}$ and a
peak of 0.57 Jy.  A two-dimensional gaussian fit to the peak channel
yields a position of (B1950) R.A.  18$^{\rm h}$ 16$^{\rm m}$
13.077$^{\rm s}$, Dec $-20^\circ$ 48$'$ 45.46$''$ or (J2000) R.A.  
18$^{\rm h}$ 19$^{\rm m}$ 12.170$^{\rm s}$, Dec $-20^\circ$ 47$'$ 28.09$''$.

There have been significant changes in the velocity structure of the
masers during the period 1995-99.  First, we note that the 1~Jy
feature at +3~km~s$^{-1}$ from the 100-m observations in 1995 was {\it
not} detected by our VLA observations in 1999.  Because of the low
angular resolution of the 100-m data, no position information is
available for the +3~km~s$^{-1}$ maser.  It seems plausible that it 
was near the position of the newly detected 20~km~s$^{-1}$ maser,
rather than coincident with VLA~3, but this is not confirmed.  
The VLA-3 maser emission shows a significantly
different velocity structure between the 1998 May observations of
Mart\'\i{} et al. (1999) and our 1999 June observations.  The
Mart\'\i{} et al. spectrum (with a bandwidth from $-$90 to $-$48~\kms)
shows emission from $-$72.6 to $-$50.2~\kms, with the strongest
component at $-$61.1~\kms.  Our spectrum shows the strongest component
now at $-$86.0~\kms.  Moreover, because of our wider bandwidth, we
also detect maser emission at this position between $-$50 and
$-$37~\kms{} (see figure~3).

\subsection{ Implications of the Newly Detected Maser Emission}

Water masers have been observed numerous times in HH80-81, including
Rodr\'\i guez et al. (1978), Rodr\'\i guez et al. (1980),
G\'omez et al. (1995), and Mart\'\i{} et al. (1999).  Most of these
observations concentrated on the high velocity component at VLA-3, and
used a relatively narrow bandpass centered near $-$70~\kms.  The
velocity components that we detect in the range of $-$50 to $-40$~\kms
are clearly associated with the VLA-3 continuum component.  However,
our broadband data lack the spatial resolution to resolve the distinct
spectral components.  Mart\'\i{} et al. 1999 resolved the VLA-3 maser
emission into two clusters of components, one of which shows a linear
structure.  The intermediate velocity components that we detect might
lend further insight into the nature of these two clumps; high angular
resolution observations of these velocity components would be
worthwhile.  Although water masers are well-known to be variable, the
differences between the published spectra of the VLA-3 masers are
substantial.  This strong variability is almost certainly relevant to 
the physical interpretation of these masers.

The positive velocity maser, at +20~\kms, may prove useful for our
understanding of the thermal jet associated with HH80-81.  The maser
is clearly aligned with the axis of the thermal jet (figure 5).  The
location of the maser at the extremum of the jet continuum emission
may indicate that the maser occurs at an interface region where the
jet interacts with its surroundings.  This interpretation is supported
by the fact that the maser occurs at the edge of a molecular clump,
seen in ammonia emission (G\'omez et al. 2003).  We note, however,
that the maser is offset from IRS~2 by $\sim$~1\farcs 5. (Stecklum et
al. 1997).  Radiative pumping mechanisms in dusty media have been
proposed for water masers (e.g., Babkovskaia \& Poutanen 2004).  Thus,
the maser may owe its existence to a nearby radiation source rather
than to the interaction of the thermal jet with the ambient cloud
material.

\section{Summary}

We present single-dish water maser observations toward 33 galactic
ultracompact HII regions.  We detect maser emission in 18 of these,
two of which (G11.11$-$0.40 and G18.30$-$0.39) are new detections.
Significant maser variability is noted for some sources.  No
correlations were found between various infrared and radio
indicators and the presence, absence, strength, or velocity spread of
maser emission.  We do find a weak trend for increasing maser
luminosity with increasing far infrared luminosity, in accord with
trends previously reported in the literature.  We compare the
detection rate of this survey (55\%) with those of other surveys, and
find that the differences can be explained in terms of differing
sensitivities and source selection criteria.

We also present interferometric water maser observations toward one
source in our sample (G10.84$-$2.59, HH80-81, GGD 27) which presented
an unusually broad range of velocity components.  We report previously
undetected, intermediate velocity maser emission, coincident with the
source VLA-3.  We also report the first detection of positive velocity
maser emission at the tip of the thermal jet.

\acknowledgments

We thank F. Wyrowski for helpful discussions and 
L. F. Rodr\'\i quez and Y. G\'omez for
providing the 3.6~cm continuum and the ammonia maps of HH80-81.
We particularly thank B. Clark for scheduling {\it ad hoc} VLA
observations of the HH80-81 region.
S.K. acknowledges financial support from DGAPA, UNAM project 118401 and
Project 36568-E, CONACyT, M\'exico.
P.H. acknowledges financial support from
NSF grant AST-0454665 and Research Corporation grant CC4996.
This research has made use of the SIMBAD database,
operated at CDS, Strasbourg, France.

\clearpage

\begin{figure}
\figurenum{1}
\plotone{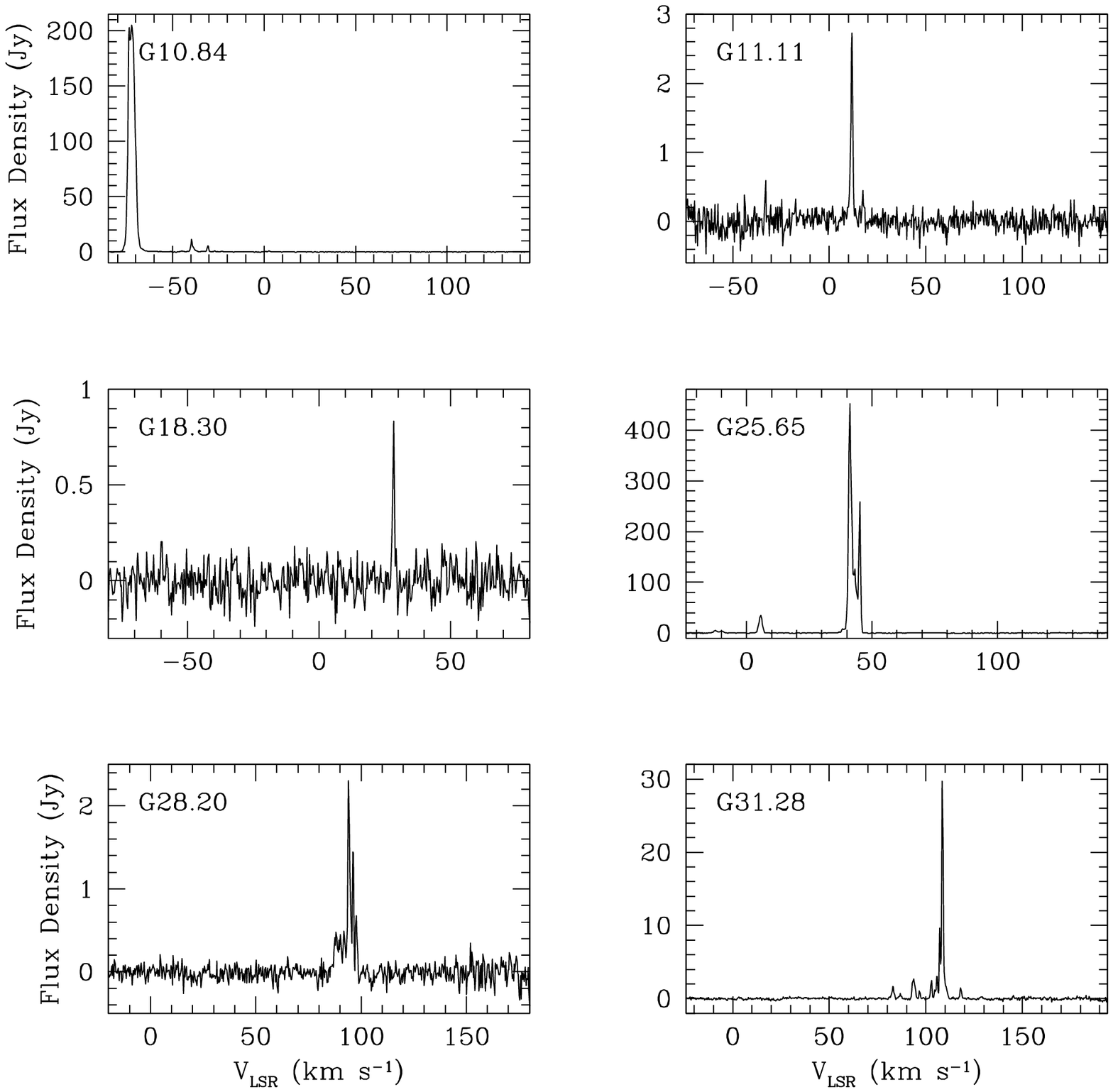}
\caption{Single-dish spectra. Spectra were obtained with the 100~m telescope
in 1995 September.  The spectral resolution in all plots is 0.33~km~s$^{-1}$.
The median $1\sigma$ {\it rms} noise level is 0.11~Jy.}
\end{figure}

\clearpage
\begin{figure}
\figurenum{1 continued}
\plotone{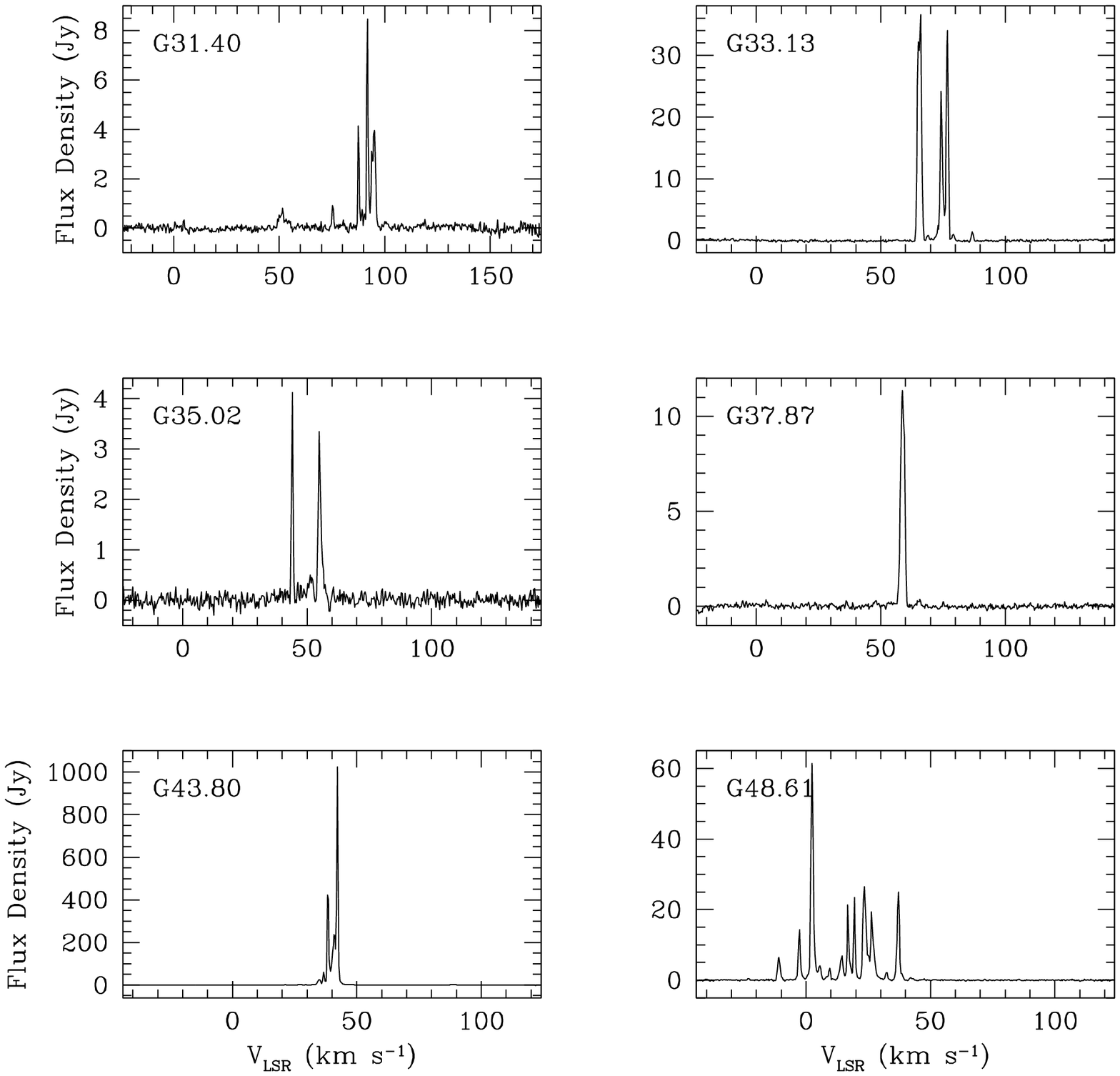}
\caption{}
\end{figure}

\clearpage
\begin{figure}
\figurenum{1 continued}
\plotone{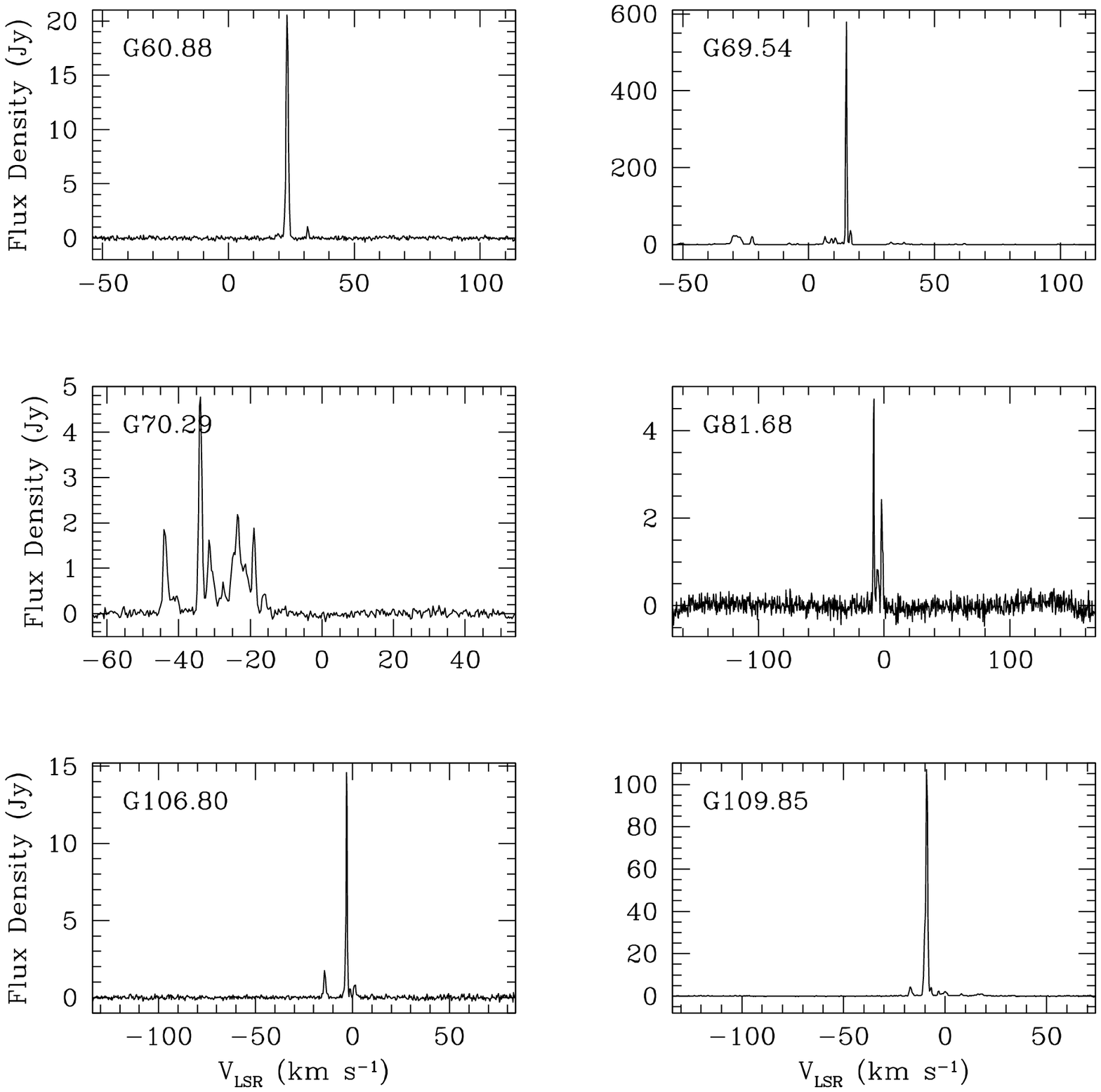}
\caption{}
\end{figure}

\clearpage
\begin{figure}
\figurenum{2}
\plotone{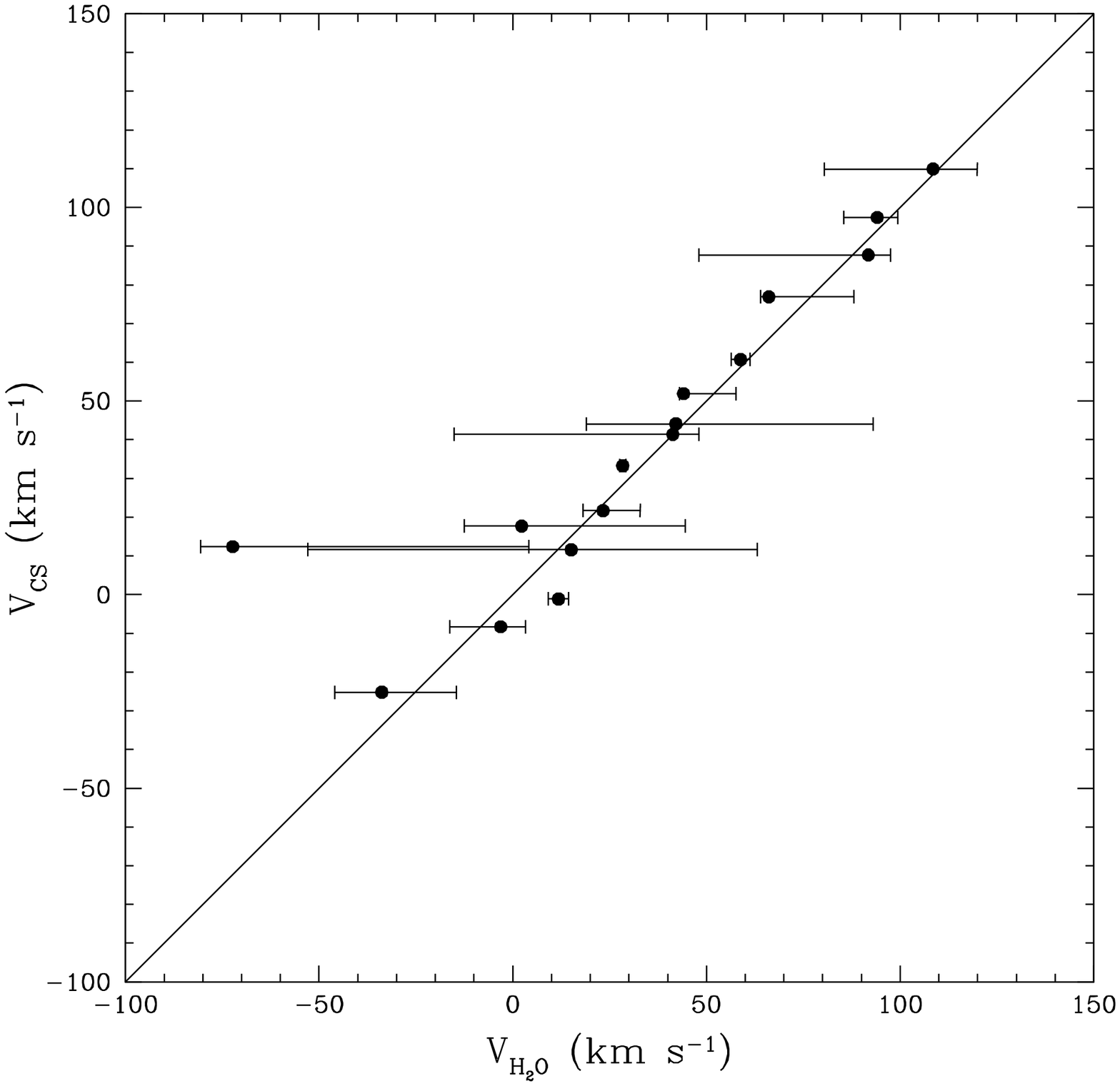}
\caption{Water maser velocities compared to host molecular cloud CS velocities
from Bronfman et al. (1996).  The points show the peak maser component velocity
(and the CS velocity) while the horizontal bars show the full range of maser 
velocity components.}
\end{figure}

\clearpage
\begin{figure}
\figurenum{3}
\plotone{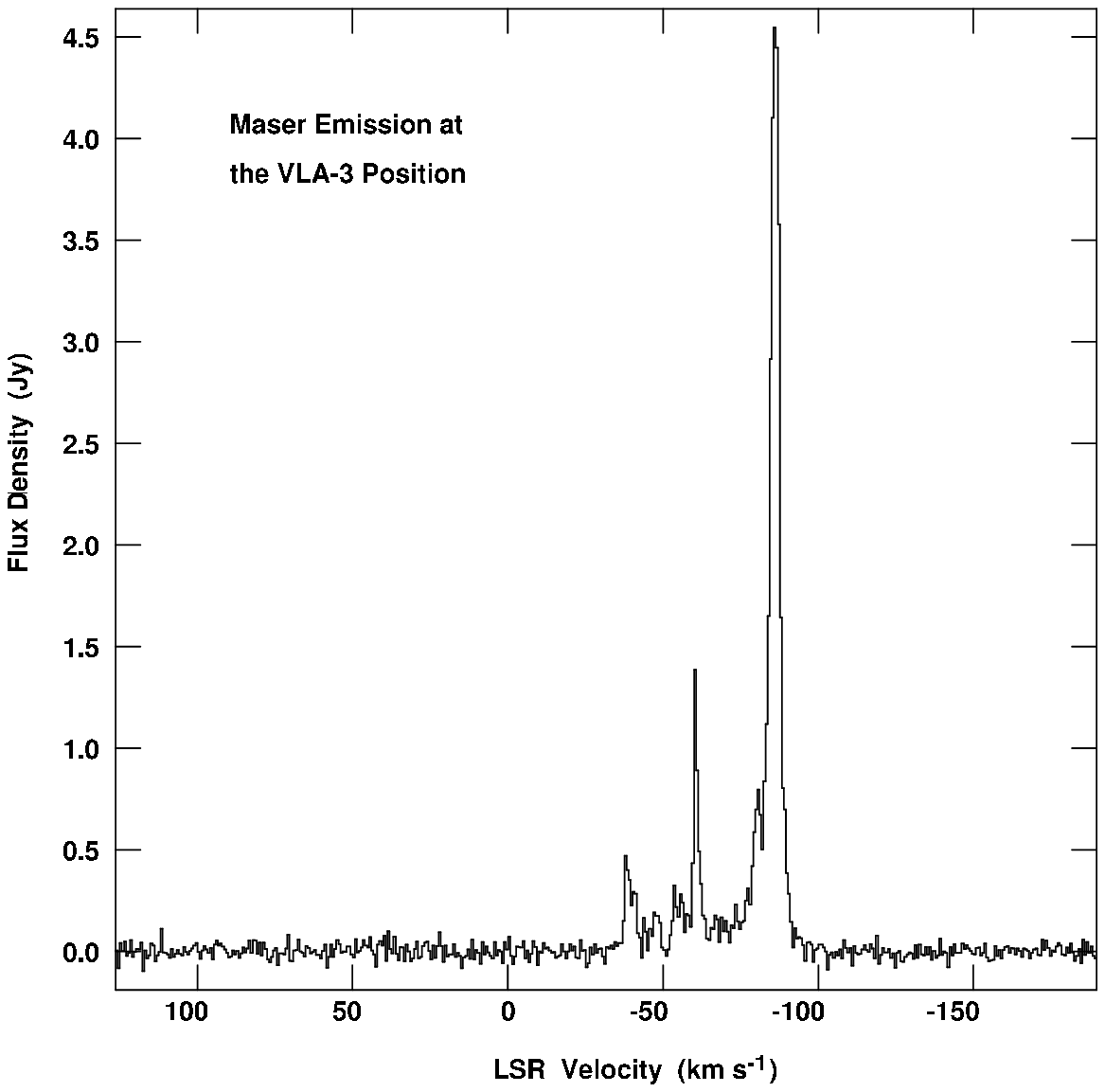}
\caption{Broad-band maser spectrum at the position of VLA-3.}
\end{figure}

\clearpage
\begin{figure}
\figurenum{4}
\plotone{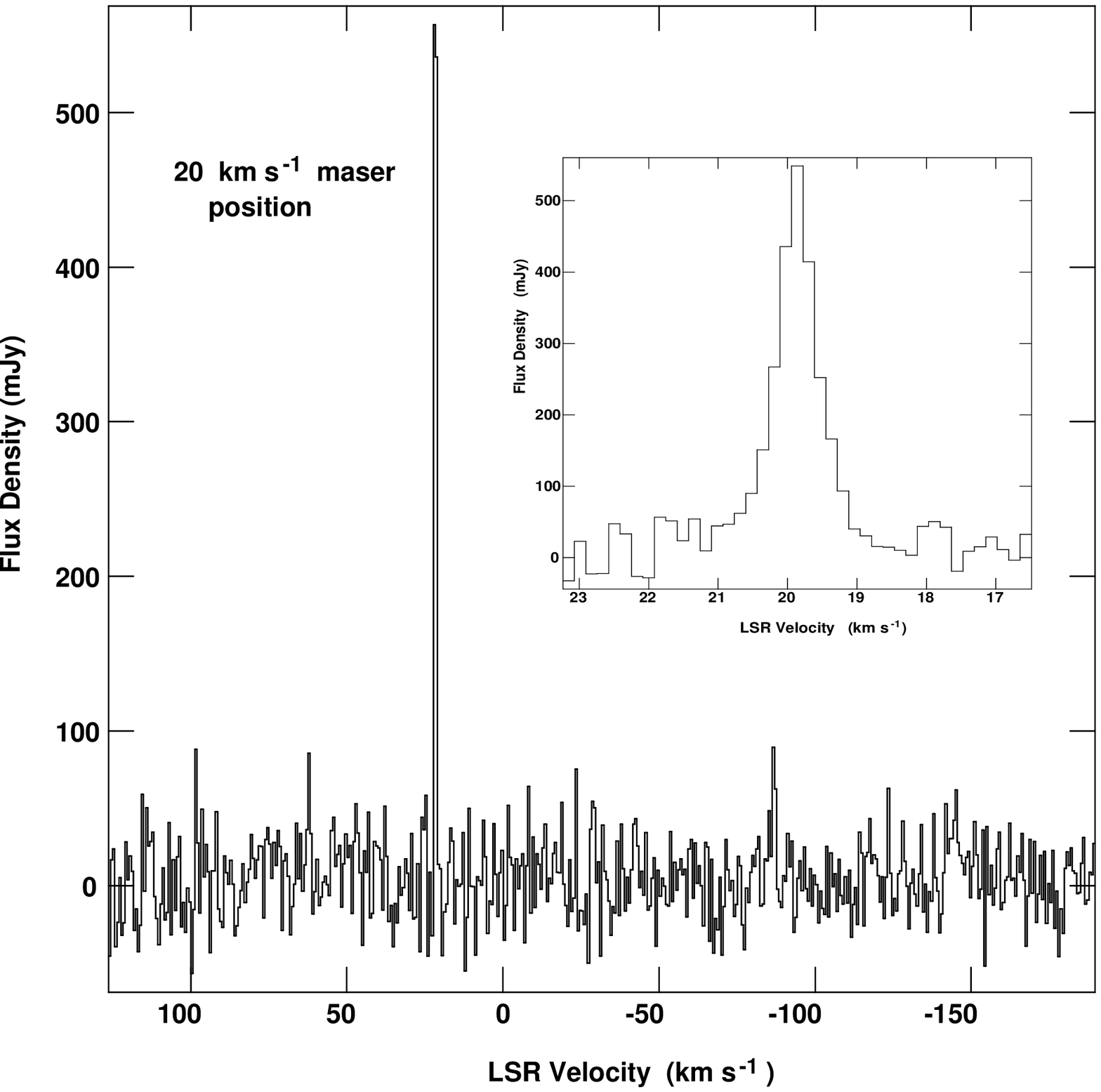}
\caption{Broad-band and narrow-band (inset) spectra of the 20~km~s$^{-1}$ 
maser aligned with the thermal jet of G10.84$-$2.59 (HH80-81).  
The broad-band spectrum was
obtained in 1999 June with angular resolution of 4\farcs 5 $\times$ 2\farcs 3
and spectral resolution of 0.66~km~s$^{-1}$.  The narrow-band spectrum (inset)
was obtained in 1999 November with  0\farcs 52 $\times$ 0\farcs 26 angular
resolution and 0.16~km~s$^{-1}$ spectral resolution.
}
\end{figure}

\clearpage
\begin{figure}
\figurenum{5}
\plotone{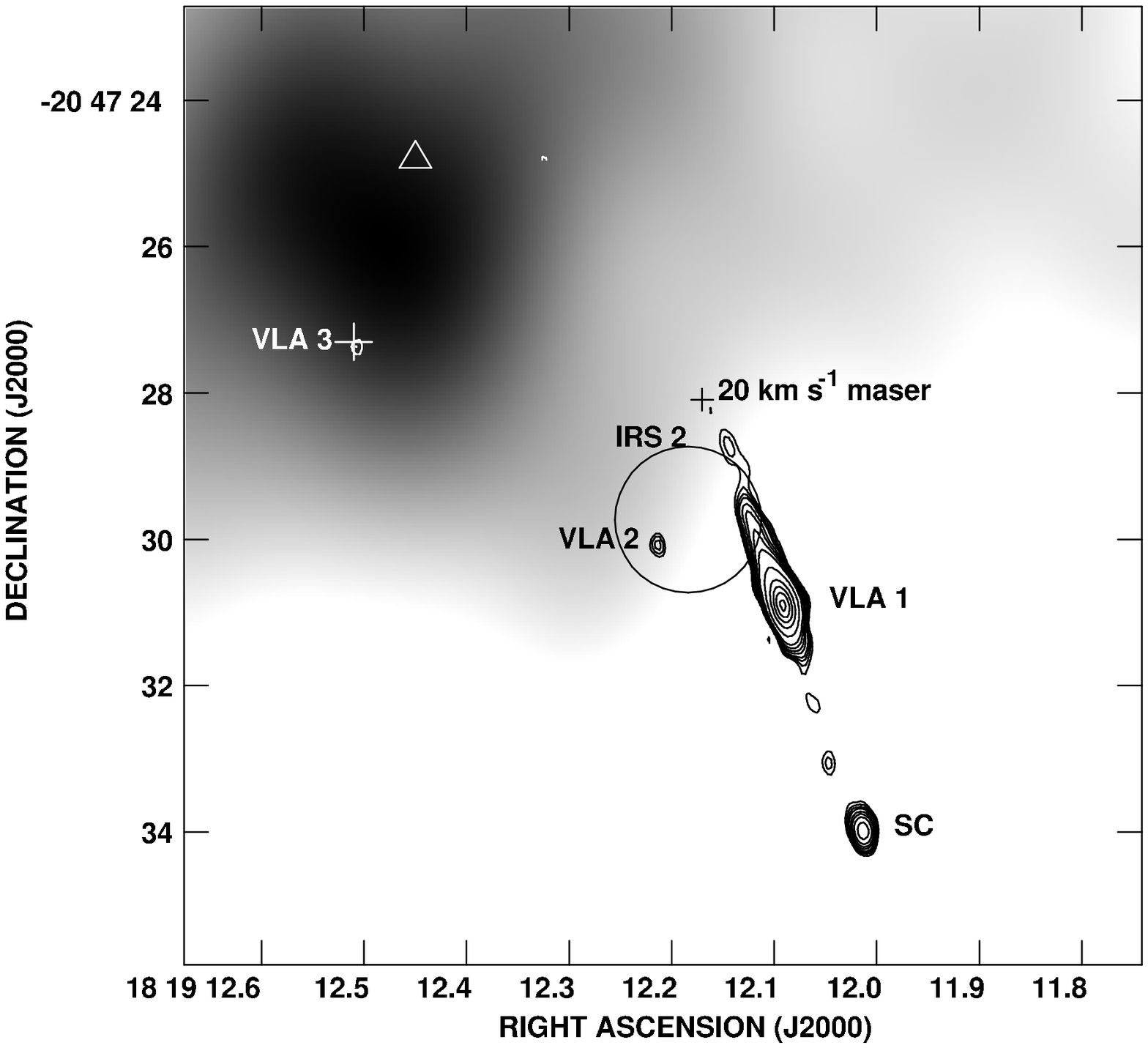}
\caption{Source positions in the HH80-81 field.
Contours show 3.6~cm continuum emission (G\'omez et al. 1995) while
greyscale shows NH$_3$ (1,1) emission (G\'omez et al. 2003).
The 20 km s$^{-1}$ water maser is seen aligned with the thermal jet, at
the border of the molecular gas. The other site of water maser emission, 
VLA-3, is associated with much denser molecular gas. The triangle shows
the position of a 44~GHz methanol maser (Kurtz, Hofner \& Vargas \'Alvarez,
2004) while the circle shows the approximate size and position of IRS2
(Stecklum et al. 1997).}
\end{figure}

\clearpage

\begin{deluxetable}{lrrrrc}
\tabletypesize{\scriptsize}
\tablecaption{Observed Sources\label{tbl-1}}
\tablewidth{0pt}
\tablehead{
\colhead{Source} & \colhead{$\alpha$ (B1950)} & \colhead{$\delta$ (B1950)} &
\colhead{$\Delta \,$V}  & \colhead{ $3\,\sigma$} & \colhead{Maser} \\
& \colhead{(h m s)} & \colhead{($\,^{\circ}\quad^{\prime}\quad^{\prime\prime}\,$)} &
\colhead{km s$^{-1}$} & \colhead{(Jy)} & \colhead{Detection}
}
\startdata
G9.88$-$0.75    & 18 07 19.3 & -20 46 22 & $-$24.3 ---  144.1 & 0.51 & N \\
G10.84$-$2.59   & 18 16 13.0 & -20 48 49 & $-$84.1 ---  144.0 & 0.44 & Y \\
G11.11$-$0.40   & 18 08 34.4 & -19 31 23 & $-$74.1 ---  143.8 & 0.64 & Y$^b$ \\
G18.15$-$0.28   & 18 22 11.6 & -13 17 23 & $-$168.5 --- 168.2 & 0.34 & N \\
G18.30$-$0.39   & 18 22 53.1 & -13 12 08 & $-$168.5 --- 168.2 & 0.52 & Y$^b$ \\
G19.49$+$0.14   & 18 23 16.6 & -11 54 20 & $-$168.5 --- 168.2 & 0.34 & N \\
G25.65$+$1.05   & 18 31 40.2 & -06 02 06 & $-$23.9 ---  143.9 & 0.50 & Y \\
G28.20$-$0.05   & 18 40 19.4 & -04 17 06 & $-$24.1 ---  183.9 & 0.69 & Y \\
G28.29$-$0.36   & 18 41 36.3 & -04 21 02 & $-$168.5 --- 144.2$^{a}$ & 0.26 & N \\
G31.28$-$0.06   & 18 45 36.5 & -01 29 53 & $-$24.1 ---  193.8 & 0.36 & Y \\
G31.40$-$0.26   & 18 46 57.5 & -01 32 33 & $-$24.1 ---  173.8 & 0.38 & Y \\
G33.13$-$0.09   & 18 49 34.3 & +00 04 31 & $-$24.1 ---  143.8 & 0.38 & Y \\
G35.02$+$0.35   & 18 51 29.0 & +01 57 29 & $-$24.1 ---  143.8 & 0.35 & Y \\
G35.57$+$0.07   & 18 53 29.9 & +02 18 56 & $-$168.5 --- 168.2 & 0.21 & N \\
G37.87$-$0.40   & 18 59 24.5 & +04 08 27 & $-$24.1 ---  143.8 & 0.34 & Y \\
G43.80$-$0.13   & 19 09 30.9 & +09 30 46 & $-$44.1 ---  123.8 & 0.46 & Y \\
G48.61$+$0.02   & 19 18 12.9 & +13 49 43 & $-$44.1 ---  123.8 & 0.37 & Y \\
G53.61$+$0.05   & 19 28 05.6 & +18 13 43 & $-$168.5 --- 168.2 & 0.23 & N \\
G60.88$-$0.13   & 19 44 13.7 & +24 28 05 & $-$54.1 ---  113.8 & 0.39 & Y \\
G69.54$-$0.98   & 20 08 09.8 & +31 22 41 & $-$54.1 ---  113.8 & 0.42 & Y \\
G70.29$+$1.60   & 19 59 50.1 & +33 24 19 & $-$64.1 ---   53.7 & 0.20 & Y \\
G76.18$+$0.13   & 20 21 53.7 & +37 28 48 & $-$168.5 --- 168.2 & 0.26 & N \\
G77.97$-$0.01   & 20 27 45.8 & +38 51 18 & $-$168.5 --- 168.2 & 0.18 & N \\
G78.44$+$2.66   & 20 17 53.0 & +40 47 07 & $-$168.5 --- 168.2 & 0.23 & N \\
G79.30$+$0.28   & 20 30 40.2 & +40 05 49 & $-$168.5 --- 168.2 & 0.23 & N \\
G79.32$+$1.31   & 20 26 22.6 & +40 43 37 & $-$168.5 --- 168.2 & 0.21 & N \\
G80.87$+$0.42   & 20 35 04.4 & +41 25 54 & $-$168.5 --- 168.2 & 0.23 & N \\
G81.68$+$0.54   & 20 37 14.2 & +42 09 15 & $-$168.5 --- 168.2 & 0.30 & Y \\
G106.80$+$5.31  & 22 17 40.3 & +63 03 36 & $-$134.1 ---  83.8 & 0.38 & Y \\
G109.85$+$2.10  & 22 54 13.5 & +61 44 46 & $-$134.1 ---  73.9 & 0.34 & Y \\
G110.21$+$2.63  & 22 55 03.0 & +62 22 25 & $-$168.5 --- 168.2 & 0.21 & N \\
G111.28$-$0.66  & 23 13 52.6 & +59 45 38 & $-$228.5 --- 108.2 & 0.26 & N \\
G111.61$+$0.37  & 23 13 21.3 & +60 50 51 & $-$128.5 --- 108.2 & 0.23 & N \\
\enddata
\tablenotetext{a}{No data between $-$35.9 and $-$24.3 km s$^{-1}$.}
\tablenotetext{b}{New detection, not previously reported in the literature.}
\end{deluxetable}
\clearpage

\begin{deluxetable}{lrrrrr}
\tabletypesize{\scriptsize}
\tablecaption{Observed Maser Parameters\label{tbl-2}}
\tablewidth{0pt}
\tablehead{
 & \colhead{ S$_{\rm max}$} & \colhead{V$_{\rm max}$} & \colhead{$\Delta \,$V} & 
\colhead{$\int$S dV } \\
\colhead{Source} & \colhead{(Jy)} & \colhead{km s$^{-1}$} & \colhead{km s$^{-1}$}
 & \colhead{(Jy km s$^{-1}$)}
}
\startdata
G10.84--2.59    & 205.1 & $-$72.3 & $-$80.6 - 4.2  & 954.2 \\
G11.11--0.40    &   2.7 & +11.8   & 5.2$^{a}$      &    3.7 \\
G18.30--0.39    &   0.8 & +28.4   & 1.6$^{a}$      &    0.7 \\
G25.65$+$1.05   & 452.3 & +41.3   & $-$15.2 - 48.0 & 1142.4 \\
G28.20$-$0.05   &   2.3 & +94.1   & 85.4 - 99.4    & 6.5 \\
G31.28$-$0.06   &  29.8 & +108.5  & 80.5 - 119.9   & 53.1 \\
G31.40$-$0.26   &   8.4 & +91.8   & 48.1 - 97.5    & 24.1 \\
G33.13$-$0.09   &  36.6 & +66.1   & 64.0 - 88.0    & 138.2 \\
G35.02$+$0.35   &   4.1 & +44.1   & 43.0 - 57.6    & 9.4 \\
G37.87$-$0.40   &  11.4 & +58.8   & 4.8$^{a}$      & 22.5 \\
G43.80$-$0.13   &1022.1 & +42.1   & 19.0 - 93.0    & 1574.1 \\
G48.61$+$0.02   &  61.2 & +2.3    & $-$12.5 - 44.5 & 272.2 \\
G60.88$-$0.13   &  20.6 & +23.3   & 18.1 - 32.9    & 24.4 \\
G69.54$-$0.98   & 579.8 & +15.1   & $-$52.9 - 63.1 & 622.7 \\
G70.29$+$1.60   &   4.8 & $-$33.8 & $-$45.9 - $-$14.5 & 23.1  \\
G81.68$+$0.54   &   4.7 & $-$8.2  & $-$10.7 - 0.7  & 9.8 \\
G106.80$+$5.31  &  14.5 & $-$3.1  & $-$16.3 - 3.3  & 17.5 \\
G109.85$+$2.10  & 107.4 & $-$9.0  & $-$19.2 - 20.4 & 162.5 \\
\enddata
\tablenotetext{a}{Single feature full width zero intensity.}
\end{deluxetable}
\clearpage

\end{document}